\documentstyle[12pt]{article}
\begin{document}

\makeatletter
\def\@maketitle{\newpage
 \null
 {\normalsize \tt \begin{flushright} 
  \begin{tabular}[t]{l} \@date 
  \end{tabular}
 \end{flushright}}
 \begin{center}
 \vskip 2em
 {\LARGE \@title \par} \vskip 1.5em {\large \lineskip .5em
 \begin{tabular}[t]{c}\@author 
 \end{tabular}\par} 
 \end{center}
 \par
 \vskip 1.5em} 
\makeatother
\topmargin=-1cm
\oddsidemargin=-.0cm
\evensidemargin=-.0cm
\textwidth=15.5cm
\textheight=22cm
\setlength{\baselineskip}{16pt}
\title{Loop Equations for + and $-$ Loops \\
              in $c = \frac{1}{2}$ Non-Critical String Theory }
\author{         
        Ryuichi NAKAYAMA \thanks{
                          nakayama@phys.hokudai.ac.jp} 
      and Toshiya SUZUKI \thanks{
                          tsuzuki@particle.phys.hokudai.ac.jp}
\\[1cm]
{\small
    Department of Physics, Faculty of Science,} \\
{\small
           Hokkaido University, Sapporo 060, Japan}
}
%
\date{
  EPHOU-95-006  \\
  hep-th/9604073 \\
  March 1996 
}

 
\maketitle

\begin{abstract}

New loop equations for all genera in $c = \frac{1}{2}$ non-critical 
string theory are constructed.   Our loop equations include two types 
of loops, loops with all Ising spins up (+ loops) and those with all 
spins down ( $-$ loops). The loop equations generate an algebra which is a 
certain extension of $W_3$ algebra and are equivalent to the $W_3$ 
constraints derived before in the matrix-model formulation of 2d 
gravity. Application of these loop equations to construction of   
Hamiltonian for $c = \frac{1}{2}$ string field theory is considered. 
 
\end{abstract}

\newpage
\newcommand {\cleqn}{\setcounter{equation}{0}}
\renewcommand {\theequation}{\thesection.\arabic{equation}}
\newcommand {\beq}{\begin{equation}}
\newcommand {\eeq}{\end{equation}}
\newcommand {\beqa}{\begin{eqnarray}}
\newcommand {\eeqa} {\end{eqnarray}}

\section{Introduction}
\cleqn
\hspace{5mm}
Loop equations frequently appear in 2d quantum gravity. 
For example, they were used to show that one-matrix model has 
various critical points that correspond to non-unitary conformal 
fields coupled to 2d gravity. \cite{kazakov}  In the case of (p,q) 
conformal fields coupled to 2d gravity it was shown that loop 
equations can be written in the form of $W_p$ constraints on 
generating function of correlation functions for scaling 
operators.\cite{fkn}\cite{dvv} Loop equations in other models such 
as ADE models, O(n) models on a random lattice have also been 
studied.\cite{kostovnokori} 

String field theory (SFT)\cite{kaku} is important for 
nonperturvative study of string theories.  Several years ago a new 
type of SFT was considered in \cite{ik} for $c=0$ noncritical 
string (2d pure gravity). Much effort has been devoted to extending 
this formalism to other $c \leq 1$ strings.\cite{ik2}$\sim$\cite{sy} 
Satisfactory formalism, however, seems to be still elusive.  

In the study of SFT it was realized that hamiltonian in SFT is in 
close relationship with loop equations.\cite{iikmns}  Therefore 
investigation of loop equations is an important step toward 
construction of SFT. In this paper we will concentrate on $c= 1/2$ 
string, {\em i.e.} continuum 
limit of Ising model coupled to 2d gravity.  In this theory many types of 
boundary conditions for Ising spins can be considered. Apparently it seems 
that by restricting the spin configurations on the loops differently we obtain 
different versions of SFT.\cite{ik2}\cite{iikmns}\cite{ns}\cite{sy}
Here we will consider only two types of spin 
configurations on loops; all Ising spins on loops are either up or down.  
Throughout this paper we will call a loop with all spins on it up a + loop, 
and a loop with all spins down a $-$ loop.  The main purpose of this paper is 
to construct loop equations in $c= 1/2$ string for + and $-$ loops.  
Such loop equations should be constructed in such a way that they are 
equivalent to $W_3$ constraints \cite{fkn}\cite{dvv}\cite{gava} derived 
in the matrix model formulation \cite{kkm} $\sim$  \cite{matrix} of 2d gravity.

$c= 1/2$ SFT for the above boundary conditions has already been presented in 
\cite{ik2}. In this reference they assumed a certain Hamiltonian for $c= 1/2$
SFT from the outset and derived loop equations as Schwinger-Dyson equations 
(SDE) in SFT.  Their loop equations, however, turned out to be a set of two 
decoupled Virasoro constraints and the connection of their loop equations with
$W_3$ constraints is not clear.  The purpose of the present work is to derive 
loop equations which are directly related to $W_3$ constraints. 

The loop operators $w_+ (l)$, $w_- (l)$ which create + and $-$ loops of length
 $l$ can be formally expanded in terms of the scaling operators.
\cite{matrix}\cite{bdss}  Therefore we can expect that if
we consider only one type of loops ({\em e.g.} + loops), the structure of 
loop equations will be very similar to that of $W_3$ constraints in matrix 
models.  Indeed loop equations for + loops can be derived in this way.  
It turns out that two independent source functions $J_+^{(1)}(l)$,
$J_+^{(2)}(l)$ have to be introduced for a + loop of length $l$. 
However, it is rather intricate to include $-$ loops into the loop equations.
We will do this by using the relationship between $w_+(l)$ and $w_-(l)$ 
through analytic continuation in $l$.  We will show that in this procedure 
we have to carefully treat the singular terms in two-loop amplitudes.
As we will see later the structure of the resultant loop equations is very 
intriguing.

In sec 2 we will write down the loop equations for + loops and show these 
equations satisfy consistency conditions.  The differential operators
in the loop equations generate the continuum version of $W_3$ algebra.
In sec 3 we derive a relationship between + and $-$ loops in terms of analytic
continuation in the length variable $l$ of loops.  We point out the problem 
which arises in Laplace transformation of this relation and derive 
correct formulae. By using these formulae the loop equations for + and $-$ 
loops are obtained in sec 4.  In sec 5 we will show that these loop equations
satisfy the consistency conditions.  These loop equations generate
a generalization of $W_3$ algebra.  An interesting structure of 
the loop equations ({\em e.g.} $T_+ - V_-$ followed by Bogoliubov 
transformation) is noticed.  In sec 6 we will consider SFT as an application 
of our loop equations.  The problem of consistency condition on string 
field Hamiltonian is discussed.   
In sec 7 we will give a brief summary and discussions. 
A reduced version of the loop equations is also derived by eliminating extra 
degrees of freedom $\bar{J}_{\pm}= (J_{\pm}^{(1)}-J_{\pm}^{(2)})/2$.
In appendix A one of the loop equations $lU_+(l) Z_{+-}=0$ obtained in 
sec 4 is shown to be equivalent to  the Virasoro constraint in matrix models.
In appendix B an explicit form of the $W_3$ current $X_+(l)$ defined in sec 5 
is presented and the algebra that $U_{\pm}$ and $X_{\pm}$ generate is 
displayed.
Some of the results in this paper were presented in \cite{ns2}.

\section{Loop Equations for + Loops}
\cleqn
\hspace{5mm}
A generating function for Green functions of scaling operators
${\cal O}_n$  $(n \neq 0 \pmod{3})$ in $c= 1/2$ string theory, 
\beqa
\tau (\mu)& = & \tau ( \mu_1, \mu_2, \mu_4, \mu_5, \mu_7, \mu_8, \cdots) 
\nonumber \\
& = & < \exp \{ \sum_{n=0}^{\infty} (\mu_{3n+1} {\cal O}_{3n+1} + \mu_{3n+2}
{\cal O}_{3n+2}) \} > 
\eeqa
satisfies the $W_3$ constraints
\beqa
L_n \tau (\mu ) & = & 0 \qquad n = -1, 0, 1, \cdots  \label{v}\\
W_n \tau (\mu) & = &  0 \qquad n = -2, -1 , 0, 1, \cdots ,
\label{w}
\eeqa
where $L_n$ and $W_n$ are differential operators with respect to $\mu$'s, 
\footnote{See eqs(\ref{tw} ),(\ref{phi} ) below}
which generate the $W_3$ algebra \cite{zamo}
\beq
[ L_n, L_m ]  =  (n - m) L_{n+m} + \frac{1}{12} n ( n^2-1) \delta_{n,-m}, 
\eeq
\beq
[ L_n, W_m ]  =  (2 n - m) W_{n+m},
\eeq
\beqa
[ W_n, W_m ] & = & -9 (n-m) U_{n+m} - \frac{1}{10} n (n^2 - 1)(n^2-4) 
\delta_{n,-m} \nonumber \\
& & + (n-m) \{ \frac{3}{2} (n^2 + 4 nm +m^2) + 
\frac{27}{2}(n+m)+21 \} L_{n+m}. \nonumber \\
& & (U_n  = \sum_{k \leq -2} L_k L_{n-k} + \sum_{k \geq -1 } L_{n-k} L_k) 
\eeqa
These results for $c=1/2$ string were first conjectured in the study of SDE 
for large-N matrix models \cite{fkn} \cite{dvv} and confirmed in \cite{gava}.
These $L_n$'s and $W_n$'s can be succinctly expressed in terms of a complex 
$Z_3$-twisted scalar field $\phi (z)$ 
\beqa
T(z) & = & - : \partial_z \phi^{*} (z) \partial_z \phi (z) : + \frac{1}{9
z^2} \equiv \sum_n z^{-n-2} L_n, \nonumber \\
W(z) & = & (\partial_z \phi (z))^3 + (\partial_z \phi^{*}(z))^3 \equiv \sum_n 
z^{-n-3} W_n,
\label{tw}
\eeqa
where the fields $\phi (z)$, $\phi ^{*}(z)$ have the following mode expansions
\footnote{Normal ordering $: \cdots :$ is defined by regarding $\mu$ as an 
annihilation operator and $\partial /\partial \mu$ a creation operator.}
\beqa
\phi (z) & = & \sum_{n=0}^{\infty}  ( z^{n+ \frac{1}{3}} \frac{\mu_{3 n +1}}
{\sqrt{g}} - z^{-n-\frac{2}{3}} \frac{3 \sqrt{g}}{3n+2} 
\frac{\partial}{\partial \mu_{3n+2}}), 
\nonumber \\
\phi^{*} (z) & = & \sum_{n=0}^{\infty}  (- z^{n+ \frac{2}{3} }
\frac{\mu_{3 n +2}}{\sqrt{g}} + z^{-n-\frac{1}{3}} \frac{3\sqrt{g}}{3n+1} 
\frac{\partial}{\partial \mu_{3n+1}}). 
\label{phi}
\eeqa
Here $g$ is a string-coupling constant.

The relationship between the SDE for matrix models and the loop equations is
well known \cite{ik}\cite{ik2}\cite{iikmns}\cite{ns}.  
The functional differential operaters appearing in 
loop equations generate a `continuum' Virasoro algebra.\cite{ik2} 
From the above results we are naturally led to introduce two independent 
source functions $J_+^{(1)}(l), J_+^{(2)}(l)$ to construct loop equations 
for + loops.  
Actually it is not difficult to figure out that loop equations take the 
following forms
\beqa
l T_+(l) Z_+ [ J_+^{(1)},J_+^{(2)} ] & = & 0, \label{le+1} \\
l^2 W_+(l) Z_+ [ J_+^{(1)},J_+^{(2)} ] & = & 32 \{ 6 \delta''(l) - 
t \delta (l) \} Z_+ [ J_+^{(1)},J_+^{(2)} ]
\label{le+2}
\eeqa
Here $Z_+$ is a generating function for amplitudes of + loops and 
$T_+(l)$, $W_+(l)$ are given by 
\beq
T_+(l) = \{ D_+^{(1)} * D_+^{(2)} + g \sum_{r=1}^{2} (lJ_+^{(r)}) 
\triangleleft D_+^{(r)} \}(l), 
\eeq
\beqa
W_+(l) & = & \sum_{r=1}^{2} \{ D_+^{(r)} *D_+^{(r)} *D_+^{(r)} + 3 g
(lJ_+^{(3-r)}) \triangleleft (D_+^{(r)} *D_+^{(r)}) \nonumber \\
& & + 3 g^2 (lJ_+^{(3-r)}) \triangleleft ((lJ_+^{(3-r)}) \triangleleft 
D_+^{(r)} ) \} (l) 
\eeqa
Here $D_+^{(r)}(l)$ stands for $\delta / \delta J_+^{(r)}(l)$. 
The symbols $*$ and $\triangleleft$ represent
the following integrals.\cite{iikmns}
\beq
(f * g)(l)  \equiv  \int_0^{l} dl' f(l')g(l-l'), \qquad 
(f \triangleleft g)(l)  \equiv  \int_0^{\infty} dl' f(l')g(l'+l).
\eeq
Later we will also need the integral
\beq
(f \triangleright g)(l) \equiv \int_0^{\infty} dl f(l'+l)g(l')
\eeq

It has to be stressed that
$J_+^{(1)}$ and $J_+^{(2)}$ in the above formulae are not the source functions
for the loop of length $l$ with a fixed boundary condition (+ spins).
Only the linear combination $J_+(l) = (J_+^{(1)}(l) + J_+^{(2)}(l))/2$ gives 
the source function for such a loop.  $J_+$ is decomposed into $J_+^{(1)}$ 
and $J_+^{(2)}$ according to the powers of $l$ contained.
\beq
J_+^{(1)}(l) \sim l^{-1/3 -n}, \quad J_+^{(2)}(l) \sim l^{-2/3 -n} \quad
(n \in {\bf Z})
\eeq
For example, the disk amplitudes 
\beq
f_+^{(r)} (l) = D_+^{(r)}(l) \ln Z |_{J=0, g=0}
\eeq
are given \cite{kostov} in laplace transformed forms 
\footnote{ Henceforth $^{\sim}$ will denote laplace transform} by
\beqa
{\tilde f}_+^{(1)}(\zeta) & = & \int_0^{\infty} dl e^{-\zeta l} f_+^{(1)}(l)
= (\zeta - \sqrt{\zeta^2-t})^{4/3}, \nonumber \\
{\tilde f}_+^{(2)}(\zeta) & = & \int_0^{\infty} dl e^{-\zeta l} f_+^{(2)}(l) 
= (\zeta + \sqrt{\zeta^2-t})^{4/3}
\label{disk}
\eeqa
Here $t$ is a cosmological constant.  

It is not difficult to show that the disk amplitudes (\ref{disk}) satisfy 
the loop equations (\ref{le+1}),(\ref{le+2}).  Conversely the loop equations 
determine the disk amplitudes completely. To show that these are actually 
equivalent to the $W_3$ constraints (\ref{v}),(\ref{w}) we have to factor 
out $Z_+$ into a singular piece $Z_{+}^{sing}$ and a regular
piece $Z_{+}^{ reg}$ as in the $c=0$ case. \cite{ik} 
We can show after a certain amount of calculation that 
\beq
\tau(\mu)  =  Z_{+}^{reg}[J_+^{(1)}(l) + 
                             c_1 t l^{-4/3} + c_2 l^{-7/3},J_+^{(2)}(l)]  
\eeq
with $c_1,c_2$ some constants satisfies $W_3$ constraints (\ref{v}),(\ref{w}). 

Before we consider $-$ loops, we will comment on the algebra of 
$T_+(l)$ and $W_+(l)$.  By explicit calculation we can show that these generate
the `continuum' $W_3$ algebra
\beq
[ T_+(l), T_+(l') ] = g (l - l') T_+ (l + l'),
\label{cm1}
\eeq
\beq
[ T_+(l), W_+(l') ] = g (2 l - l') W(l + l'),
\label{cm2}
\eeq
\beqa
[ W_+(l), W_+(l') ] & = & 9 g (l-l') (T_+*T_+)(l+l') + 18g (l-l') (V_+ 
\triangleleft T_+)(l+l') \nonumber \\
& & - \frac{3}{2} g^2 (l-l') (l^2 + 4 ll' +l'^2) T_+(l+l'). 
\label{cm3}
\eeqa
Here $V_+(l)$ is defined by
\beq
V_+(l) = \{ g \sum_{r=1}^{2} (lJ_+^{(r)}) \triangleright D_+^{(r)}
     + g^2 (lJ_+^{(1)}) * (lJ_+^{(2)}) \}(l)
\eeq
and corresponds to Virasoro generators $L_n$ with $n \leq -2$, while
$T_+(l)$ to $L_n$ with $n \geq -1$: we have 
\beq
[ T_+(l), V_+(l') ]  =  g (l+l') T_+(l-l') \theta (l-l')
      + g (l + l') V_+(l'-l) \theta (l'-l), 
\label{cm4}
\eeq
\beq
[ V_+(l), V_+(l') ]  =  - g (l-l') V_+(l+l'),
\label{cm5}
\eeq
where $\theta$ is a step function.

The algebra (\ref{cm1}) $\sim$ (\ref{cm3}) is not strictly closed because $V_+$
appears on the right-hand side of (\ref{cm3}).  Nonetheless this is 
sufficient for consistency of the loop equations (\ref{le+1}), (\ref{le+2}). 
\footnote{
Strictly speaking, for the proof of consistency we need more information 
than (\ref{le+1}).  Actually by explicitly solving (\ref{le+1}) and 
(\ref{le+2}) for $Z_+$ as perturbation series in $g$ we can show 
that $T_+(l)Z_+ = - g \delta(l) \frac{\partial}{\partial t} Z_+ $.  
A similar equation was also noticed in \cite{sy}
}
We can also show as in \cite{ik2} that the right-hand side of (\ref{le+2}) does
not make (\ref{le+1}),(\ref{le+2}) inconsistent.

\section{Relation between Loop Amplitudes for + Loops and  $-$ Loops}
\cleqn
\hspace{5mm}
In this section we will derive formulae which relate + and $-$ loops by
analytic continuation in variables $l$, $\zeta$.  
In the next section by using these formulae we will derive loop equations 
for both types of loops, + and $-$ loops.  

Let us denote the  loop operators which create a loop of length $l$ 
with + spins and that with $-$ spins by $w_+(l)$ and $w_-(l)$, respectively. 
These are related by 
\beq
w_-(l) =  - w_+(e^{3 \pi i} l)
\label{rel1}
\eeq
This relation is implicit in the representation for loop amplitudes 
in terms of heat kernels.\cite{matrix}\cite{mss}
For example one- and two-loop amplitudes can be written as
\beq
< w_{\pm}(l) > \propto \int_t^{\infty} dx <x| e^{\pm l L} |x>, 
\eeq
\beq
< w_+(l_1) w_{\pm}(l_2) >_c \propto \int_t^{\infty} dx \int_{-\infty}^t dy <x|
               e^{l_1 L} |y> <y| e^{\pm l_2 L} |x>
\eeq
Here $L$ is a third-order Lax operator given by
\beq
L = \frac{1}{2} \{ - (\frac{d}{dx})^3 + 3 x^{1/3} \frac{d}{dx} \}
\eeq
These representations can be obtained by computing correlation functions in 
two-matrix model \cite{kaza} by using the method of orthogonal polynomials 
and taking continuum limits of them.\footnote{
Loop amplitudes in (p,q) gravity were computed in \cite{dks} }
 A $-$ loop of length $l$ appears 
as a + loop of `length $-l$' in these representations. 

The reason for an extra minus sign on the right-hand side of  (\ref{rel1}) 
can be understood if we expand $w_{\pm}(l)$ in powers of $l$.
\beqa
w_+(l) & = & \sum_{n=0}^{\infty} {\cal O}_n^{(1)} l^{n+ 1/3} + 
\sum_{n=0}^{\infty} {\cal O}_n^{(2)} l^{n+ 2/3},  \nonumber \\
w_-(l) & = & \sum_{n=0}^{\infty} {\cal O}_n^{(1)} (-1)^n l^{n+ 1/3} - 
\sum_{n=0}^{\infty} (-1)^n {\cal O}_n^{(2)} l^{n+ 2/3} 
\label{exp}
\eeqa
Here ${\cal O}_{3n+1} \equiv {\cal O}_n^{(1)}$ and ${\cal O}_{3n+2} \equiv 
{\cal O}_n^{(2)}$ are scaling operators.  ${\cal O}_{2n}^{(1)}$ and
${\cal O}_{2n+1}^{(2)}$ are $Z_2$-even operators, while  
${\cal O}_{2n+1}^{(1)}$ and ${\cal O}_{2n}^{(2)}$ are odd. 
We have  $\exp \{ 3 \pi i \}$ instead of $\exp \{ \pi i \}$
in (\ref{rel1}), because loop amplitudes have to be real functions.

To derive loop equations we need to laplace transform the relation 
(\ref{rel1}).  As far  as loop amplitudes can
be expanded in fractional powers of $l$ as in (\ref{exp}),  we can easily 
show that
\beq
\tilde{w}_-(\zeta) =  \tilde{w}_+(e^{3 \pi i} \zeta)
\label{rel2}
\eeq
This relation (\ref{rel2}) is valid for n-loop amplitudes with $n \geq 3$. 
In the case of one- and two-loop amplitudes we have to be careful in the 
treatment of singular terms.  One-loop amplitudes can be expanded in powers of 
$l$, although first few terms are singular.  Laplace transformation of these
singular terms can be performed as usual in the sense of analytic 
continuation and the relation (\ref{rel2}) is still valid.  On the other hand 
two-loop amplitudes contain the singular terms which are not expandable.
\beq
< w_+(l_1) w_{\pm}(l_2) >_{c,sing} =  \frac{\sqrt{3}}{2 \pi} g 
\frac{(l_1 l_2)^{1/3}}{l_1 \pm l_2} (l_1^{1/3} \pm l_2^{1/3})
\label{sing1}
\eeq
The laplace transforms of these are given by
\beq
< \tilde{w}_+(\zeta_1) \tilde{w}_+(\zeta_2)>_{c,sing} =
g \frac{(\zeta_1/\zeta_2)^{2/3} +2(\zeta_1/\zeta_2)^{1/3} 
+ 2 (\zeta_2/\zeta_1)^{1/3} + (\zeta_2/\zeta_1)^{2/3} - 6}
{3 (\zeta_1 - \zeta_2)^2},  
\label{sing2a}
\eeq
\beq
< \tilde{w}_+(\zeta_1) \tilde{w}_-(\zeta_2)>_{c,sing} =
g \frac{(\zeta_1/\zeta_2)^{2/3} - 2(\zeta_1/\zeta_2)^{1/3} 
- 2 (\zeta_2/\zeta_1)^{1/3} + (\zeta_2/\zeta_1)^{2/3} +3}{3 (\zeta_1 +
 \zeta_2)^2}.
\label{sing2b}
\eeq
These yield the following relation
\beq
< \tilde{w}_+(\zeta_1) \tilde{w}_-(\zeta_2)>_c = < \tilde{w}_+(\zeta_1)
\tilde{w}_+( e^{3\pi i} \zeta_2)>_c + \frac{3g}{(\zeta_1 + \zeta_2)^2}
\label{2pt}
\eeq

We should also decompose the loop operators $w_+(l)$, $w_-(l)$ according to 
the powers of $l$.  
\beq    
w_{\pm}(l) = w_{\pm}^{(1)}(l) + w_{\pm}^{(2)}(l), 
\eeq
\beq
w_{\pm}^{(1)}(l) \sim l^{1/3+n}, \quad w_{\pm}^{(2)}(l) \sim l^{2/3+n}, 
\quad (n \in {\bf  Z})
\eeq
It is natural to put the second term on the right-hand side of (\ref{2pt}) 
into the relation for $< \tilde{w}^{(1)} \tilde{w}^{(2)} >$ and 
we obtain the following relations.
\beqa
& & < \tilde{w}_+^{(r_1)}(\zeta_1) \cdots \tilde{w}_+^{(r_n)}(\zeta_n)
 \tilde{w}_-^{(s_1)}(\xi_1) \cdots \tilde{w}_-^{(s_m)}(\xi_m)>_c \nonumber \\
& & = < \tilde{w}_+^{(r_1)}(\zeta_1) \cdots \tilde{w}_+^{(r_n)}(\zeta_n)
 \tilde{w}_+^{(s_1)}(e^{3 \pi i} \xi_1) \cdots \tilde{w}_+^{(s_m)}
( e^{3 \pi i} \xi_m)>_c,
\nonumber
\eeqa
\beq
 < \tilde{w}_+^{(r)}(\zeta) \tilde{w}_-^{(s)}(\xi) >_c
 =  <\tilde{w}_+^{(r)}(\zeta) \tilde{w}_+^{(s)}( e^{3 \pi i} \xi) >_c
   + \frac{3}{2} g \frac{\delta_{r+s,3}}{(\zeta + \xi)^2}
\label{rel3}
\eeq
Here $n+m \neq 2$ and $r,s,r_i,s_i = 1,2$.
These formulae will play an important role in the next section.

\section{Inclusion of $-$ Loops into Loop Equations}
\cleqn
\hspace{5mm}
We will  derive the loop equations for loops with both types of spin 
configurations.  For that purpose we have to introduce two more source 
functions $J_-^{(1)}(l)$, $J_-^{(2)}(l)$ for $-$ loops.  We start with 
\beq
D_+^{(1)}(l_2) \frac{1}{Z_+} l_1 T_+(l_1) Z_+  =0
\label{start}
\eeq
and set $J_+^{(1)}=J_+^{(2)}=0$. By laplace transforming this 
($\int_0^{\infty}dl_1 \int_0^{\infty} dl_2 \exp \{ -\zeta_1 l_1 - \zeta_2 l_2 
\}$) and rewriting this in terms of loop operators, we obtain
\beqa
& - \frac{\partial}{\partial \zeta_1} [ & 
  < \tilde{w}_+^{(1)}(\zeta_1) \tilde{w}_+^{(2)}(\zeta_1) 
  \tilde{w}_+^{(1)}(\zeta_2)>_c
  +  < \tilde{w}_+^{(1)}(\zeta_1) \tilde{w}_+^{(1)}(\zeta_2)>_c 
     <\tilde{w}_+^{(2)}(\zeta_1)> \nonumber \\
& &  +   <\tilde{w}_+^{(1)}(\zeta_1)> 
      < \tilde{w}_+^{(2)}(\zeta_1) \tilde{w}_+^{(1)}(\zeta_2)>_c
            \nonumber \\
& & + g \frac{\partial}{\partial \zeta_2} \{ \frac{1}{\zeta_1 - \zeta_2} (
<\tilde{w}_+^{(1)}(\zeta_1)> - <\tilde{w}_+^{(1)}(\zeta_2)>) \} ] = 0.
\label{lpe}
\eeqa
Here we replace $\zeta_2$ by $ e^{3 \pi i} \zeta_2$ and use the relation 
(\ref{rel3}) to rewrite the above equation. This yields
\beqa   
& - \frac{\partial}{\partial \zeta_1} [ & 
  < \tilde{w}_+^{(1)}(\zeta_1) \tilde{w}_+^{(2)}(\zeta_1) 
  \tilde{w}_-^{(1)}(\zeta_2)>_c
  +  < \tilde{w}_+^{(1)}(\zeta_1) \tilde{w}_-^{(1)}(\zeta_2)>_c 
     <\tilde{w}_+^{(2)}(\zeta_1)> \nonumber \\
& &  +   <\tilde{w}_+^{(1)}(\zeta_1)> 
      < \tilde{w}_+^{(2)}(\zeta_1) \tilde{w}_-^{(1)}(\zeta_2)>_c
            \nonumber \\
& & + g \frac{\partial}{\partial \zeta_2} \{ \frac{1}{\zeta_1 + \zeta_2} (
 \frac{1}{2} <\tilde{w}_+^{(1)}(\zeta_1)> + <\tilde{w}_-^{(1)}(\zeta_2)>) 
\} ] = 0.
\eeqa
This equation can be furthermore rewritten as follows
\beqa
\lefteqn{ D_-^{(1)}(l_2) \frac{l_1}{Z_{+-}} [  \int_0^{l_1} dl' D_+^{(1)}(l') 
D_+^{(2)}(l_1-l') 
 - \frac{1}{2} g \int_0^{l_1} dl' l' J_-^{(1)}(l') D_+^{(1)}(l_1-l')} 
\nonumber \\
& & - g \int_0^{\infty} dl' (l_1+l') J_-^{(1)}(l'+l_1) D_-^{(1)}(l') ]
     Z_{+-} |_{J_+^{(r)}=J_-^{(r)}=0} =0.
\eeqa
Here $Z_{+-}=Z_{+-}[J_+^{(1)},J_+^{(2)},J_-^{(1)},J_-^{(2)}]$ is a generating 
function for both types of loops.

By operating $D_+^{(r)}$ on eq (\ref{start}) arbitrary times and 
repeating this procedure we finally arrive at the new loop equation
\beqa
& & l U_+(l) Z_{+-} = 0,
\label{newle1} \\
& & l U_-(l) Z_{+-} = 0,
\label{newle2}
\eeqa
where $U_{\pm}(l)$ is the following operator.
\beqa
U_{\pm}(l) & = & \{ D_{\pm}^{(1)}*D_{\pm}^{(2)}+g\sum_{r=1}^2 (lJ_{\pm}^{(r)}) 
  \triangleleft D_{\pm}^{(r)}  
 - g \sum_{r=1}^2 (lJ_{\mp}^{(r)}) \triangleright D_{\mp}^{(r)} \nonumber \\
& & +  \alpha g \sum_{r=1}^2 (lJ_{\mp}^{(r)})*D_{\pm}^{(r)} 
 + (\alpha^2 -1) g^2 (lJ_{\mp}^{(1)})*(lJ_{\mp}^{(2)}) \}(l), \nonumber \\ 
& & \qquad \qquad \qquad (\alpha  = - \frac{1}{2}) .
\label{U}
\eeqa
Eq (\ref{newle2}) is obtained from eq (\ref{newle1})
by $Z_2$ symmetry for Ising spins. Let us note that if we set 
$J_-^{(1)}=J_-^{(2)}=0$, the loop equations (\ref{newle1}), (\ref{newle2}) 
reduce to those for only + loops (\ref{le+1}). 

Although we have derived loop equation (\ref{newle1}) by using the method of
analytical continuation (\ref{rel3}), more direct proof will be desirable.
In Appendix A we will prove that loop equation (\ref{newle1}) is equivalent 
to the Virasoro constraint (\ref{v}).  We also checked that disk and cylinder 
amplitudes satisfy (\ref{newle1}). 

Let us make a remark on the value of $\alpha$.  We might formally repeat the 
procedure after (\ref{start}) without laplace trasforming the loop equations.
Because the singular parts of two-loop amplitudes (\ref{sing2a}),
(\ref{sing2b}) respect the 
relation (\ref{rel1}), this relation might be used for any amplitudes.
In this case we would again end up with the loop equations (\ref{newle1}), 
(\ref{newle2}) but the value of $\alpha$ in (\ref{U}) would then be replaced 
by $1$. This discrepancy can be resolved if we take into account the fact that 
the $\int dl$ integrals appearing in the loop equations are divergent, 
when the lengths $l$ of loops in one-loop amplitudes vanish, and we should 
not make the analytic continuation (\ref{rel1}) in such divergent integrals.  
To the contrary Laplace-transformed loop equations like (\ref{lpe}) are 
algebraic  and cause no problems.  
Hence we should put $\alpha = - \frac{1}{2}$ as above.

\section{Consistency Conditions}
\cleqn
\hspace{5mm}
In this section we will show that $U_+(l)$ and $U_-(l)$ defined in the previous
section satisfy consistency conditions.  Let us define the operators
\beq
\hat{U}_+(l) \equiv T_+(l) - V_-(l), \qquad 
\hat{U}_-(l) \equiv T_-(l) - V_+(l),
\eeq
where $T_-(l)$ and $V_-(l)$ are obtained from $T_+(l)$ and $V_+(l)$,
 respectively, by
interchanging $J_+^{(r)}$ and $D_+^{(r)}$ with $J_-^{(r)}$ and $D_-^{(r)}$.
Because $J_+^{(r)}$ and $J_-^{(r)}$ are independent functions, we can show 
by using the commutation relations (\ref{cm1}), (\ref{cm4}),(\ref{cm5})
 that $U_{\pm}(l)$ generate the algebra
\beq
[ \hat{U}_{\pm}(l_1), \hat{U}_{\pm}(l_2) ]  =  
   g (l_1 - l_2) \hat{U}_{\pm}(l_1+l_2),  
\label{con1}
\eeq
\beqa
[ \hat{U}_+(l_1), \hat{U}_-(l_2) ] & = &
- g (l_1+l_2) \hat{U}_+(l_1-l_2) \theta (l_1-l_2) \nonumber \\
& & + g (l_1+l_2) \hat{U}_-(l_2-l_1) \theta (l_2-l_1).
\label{con2}
\eeqa
These two generators can be combined into a single one: the operator 
\beq
\hat{U}(l) \equiv \hat{U}_+(l) \theta (l) - \hat{U}_-(-l) \theta(-l)
\qquad (-\infty < l < \infty)
\eeq
generates a whole `continuum' Virasoro algebra.

The functional differential operators $U_+(l)$, $U_-(l)$ which appear in loop 
equations (\ref{newle1}), (\ref{newle2}) 
turn out to be related to $\hat{U}_+(l)$ and  $\hat{U}_-(l)$ 
by a Bogoliubov transformation
\beqa
D_{\pm}^{(r)}(l) & \rightarrow & D_{\pm}^{(r)}(l) 
                   - \frac{1}{2}g l J_{\mp}^{(3-r)}(l), \nonumber \\
J_{\pm}^{(r)}(l) & \rightarrow & J_{\pm}^{(r)}(l) \qquad (r=1,2) 
\label{bogo}
\eeqa
Because this transformation does not change the algebra,
 $U_+(l)$ and $U_-(l)$ generate two {\em coupled} Virasoro algebras
(\ref{con1}),(\ref{con2}), and
consistency conditions are satisfied.  At present the meaning of this 
transformation is not clear to us.

The analysis in the preceding and present sections can be generalized to 
$W_+(l)$. We define 
\beqa 
Y_-(l) = & & \sum_{r=1}^2 [ 3 g ((lJ_-^{(r)}) \triangleright D_-^{(3-r)}) 
\triangleright D_-^{(3-r)} 
 + 3g^2 ((lJ_-^{(r)})*(lJ_-^{(r)})) \triangleright D_-^{(3-r)} \nonumber \\ 
& & + g^3 (lJ_-^{(r)})*(lJ_-^{(r)})*(lJ_-^{(r)}) ](l),
\eeqa
and construct
\beq
\hat{X}_+(l) \equiv W_+(l)  -  Y_-(l).
\eeq
By applying transformation (\ref{bogo}) on $\hat{X}_+(l)$, we obtain 
$X_+(l)$. We will present it in Appendix B explicitly. 
 Similarly we can construct $X_-(l)$.
The operator $X_{\pm}(l)$  together with $U_{\pm}$ generate 
a closed algebra.  This is also presented in Appendix B.

To summarize, loop equations for $c=1/2$ string are given by
\footnote{
We can also show that $U_{\pm}(l) Z_{+-} = -g \delta(l) \frac{\partial}{
\partial t} Z_{+-}$ by starting from the equation in footnote 4 and applying 
the same method as in sec 4 
}
\beqa
& & lU_+(l) Z_{+-} = lU_-(l) Z_{+-} = 0, \nonumber \\
& & l^2 X_+(l) Z_{+-} = l^2 X_-(l) Z_{+-} = 
32 (6\delta''(l) - t \delta(l)) Z_{+-}
\label{loopeq}
\eeqa
This is the main result of this paper.

\section{String Field Theory for $c= 1/2$ String}
\cleqn
\hspace{5mm}
The loop equations obtained in the previous section determine the loop
amplitudes completely.  
In this section we will discuss  applications of these loop equations 
to string field theory.  

Let us first consider a theory which contains only loops with + spins.
In the string field theory we consider here\cite{ik}, the generating function 
of loop amplitudes will be expressed in the form
\beq 
Z_{+}[J_+^{(1)}, J_+^{(2)}] = 
\lim_{D \rightarrow \infty} <0| e^{-D H } \exp \{ \int_0^{\infty} dl 
\sum_{r=1,2} J_+^{(r)}(l) \Psi_+^{(r)\dagger}(l) \} |0>
\label{limit1}
\eeq
The parameter $D$ may be interpreted as geodesic distance on the world sheet.
Here $\Psi_+^{(r)\dagger}(l)$ is a creation operator for a loop of length $l$
with + spins and satisfies with the corresponding
annihilation operators $\Psi_+^{(r)}(l)$ the commutation relations
\beq
[\Psi_+^{(r)}(l), \Psi_+^{(r')\dagger}(l')] = \delta_{r,r'} \delta(l-l')
\eeq
Hamiltonian $H$ is a functional of $\Psi_+^{(r)}$, 
$\Psi_+^{(r) \dagger}$ and related to the loop equations.  
To determine its form we note that the existence of the limit (\ref{limit1}) 
implies string field SDE
\beq
\lim_{D \rightarrow \infty} \frac{\partial}{\partial D} 
<0| e^{-D H } \exp \{ \int_0^{\infty} dl 
\sum_{r=1,2} J_+^{(r)}(l) \Psi_+^{(r)\dagger}(l) \} |0> =0.
\label{limit2}
\eeq
This can be rewitten as a differential equation for $Z_+$
\beq
\hat{H}[J_+^{(r)},D_+^{(r)}] Z_+ [J_+^{(1)}, J_+^{(2)}] = 0,
\label{SDE}
\eeq
where $\hat{H}$ is obtained from $H$ by replacing $\Psi_+^{(r)}(l)$ and  
$\Psi_+^{(r)\dagger}(l)$ by $J_+^{(r)}(l)$ and $D_+^{(r)}(l)$, respectively, 
and interchanging the ordering of $J$'s and $D$'s. 

In the case of pure gravity ($c=0$), $\hat{H}$ is given by\cite{ik}
\beq
\hat{H} = \int_0^{\infty} dl  J(l) \{ l T(l) -\rho(l) \},
\eeq
where
\beq
T(l) = \int_0^l D(l') D(l-l') + g \int_0 ^{\infty} dl' J(l') D(l+l')
\eeq
and $\rho(l)$ is a tadpole term.  
If the generating function $Z[J]$ satisfies the loop equation
\beq 
lT(l) Z[J] = \rho (l) Z[J]
\eeq
$Z[J]$ also satisfies SDE (\ref{SDE}).  The converse can also be proved
\cite{ik}. 

Therefore a simple generalization to $c= 1/2$ string will be to take as 
$\hat{H}$ 
\footnote{
Similar Hamiltonian is constructed by a different method in \cite{aw2}.
Main difference between our Hamiltonian and theirs is that their Hamiltonian 
does not contain terms corresponding to the singular terms of one- and
two-loop amplitudes.  } 
\beqa
& & \hat{H}  =  \int_0^{\infty} dl J_+(l) \{ l^2 W_+(l) - \rho(l) \}  
\label{ham1} \\
& & + g \int_0^{\infty} dl_1 \int_0^{\infty} dl_2 l_1 l_2 (l_1 + l_2) 
\{ a J_+(l_1) J_+(l_2) + b \bar{J}_+(l_1) \bar{J}_+(l_2) \} T_+(l_1 + l_2),
\nonumber  
\eeqa
where $a$, $b$ are some constants not yet determined and the tadpole term is 
given by
\beq
\rho(l) = 32 ( 6 \delta''(l) - t \delta(l)) .
\eeq
$J_+(l)$ and $\bar{J}_+(l)$ are defined by 
\beq
J_+ \equiv \frac{1}{2} (J_+^{(1)} + J_+^{(2)}), \quad  
\bar{J}_+ \equiv \frac{1}{2} (J_+^{(1)} - J_+^{(2)}).
\label{J}
\eeq 
The dimensions of $\Psi_+^{(r)}$, $\Psi_+^{(r) \dagger}$ are 
given by $L^{4/3}$, $L^{-7/3}$, respectively and that of $g$ is $L^{- 14/3}$,
where $L$ denotes the dimension of the length of a loop.
Hence the dimension of $\hat{H}$ is $L^{-2/3}$ and we get the result  
$[D] \sim L^{2/3}$.  However, for the reason which will be stated in the 
next few paragraphs we can not draw a definite conclusion about 
the dimension of geodesic distance $D$ in this paper.\footnote{
For discussion of the dimension of geodesic distance $D$ see \cite{ik2},
\cite{iikmns},\cite{kkmw},\cite{ajw}  }
This construction of Hamiltonian can also be extended in an obvious way 
to the theory where $-$ loops are included.
In (\ref{ham1}) $W_+$ and $T_+$ have to be replaced by $X_+$ and $U_+$,
respectively and terms with + and $-$ interchanged should be added.

By construction our $Z_+[J_+^{(1)}, J_+^{(2)}]$ satisfies loop equations
(\ref{le+1}), (\ref{le+2}) and we may stop  at this point. But if our string 
field theory is to have geometrical meaning, each interaction in the 
Hamiltonian should give proper decomposition of world sheets into propagators,
 vertices and tadpoles. In other words, the Hamiltonian should provide us 
with Feynman rules for calculation of loop amplitudes with geodesic 
distances among loops fixed.

In \cite{ik2} a consistency condition for string field Hamiltonian was 
proposed.  Let us consider a world sheet with the topology of a cylinder.
Suppose that the minimum geodesic distance of the two boundaries is $D$.  We
are able to compute an amplitude for such a geometry by sewing two transfer
matrices by a disk.  Starting from the two boundaries two loops propagate 
by splitting and disappearing and eventually two loops from the two 
boundaries meet at a point.  Let the geodesic distance between this point and 
the two boundaries be $D_1$ and $D_2$ ($D_1 + D_2=D$).  After the two loops 
meet at a point, they merge and the loop will keep splitting and disappearing 
to form a disk.
In \cite{ik2} it was proposed to require that the resulting amplitude should
not depend on how the geodesic distance $D$ is decomposed into two. 
This is a reasonable requirement if we are to compute such amplitudes 
in string field theory.  

We performed such a consistency check of our Hamiltonian (\ref{ham1}) and
found the consistency condition in the above sense does not seem to be 
satisfied.  The obstruction seems to lie in the fact that our 
loop equations contain extra degrees of freedom $\bar{J}_{\pm}= 
(J_{\pm}^{(1)}- J_{\pm}^{(2)})/2$ in addition to $J_{\pm}=(J_{\pm}^{(1)}
+ J_{\pm}^{(2)})/2$. In the next section we will construct loop equations 
for $J_{\pm}$ only.

\section{Discussions}
\cleqn
\hspace{5mm}
In this paper we derived the loop equations (\ref{loopeq}) in $c= \frac{1}{2}$ 
string, where Ising spins on the loops are all either up or down.  
These loop equations are equivalent to the $W_3$ constraints 
\cite{fkn}\cite{dvv}\cite{gava} which were derived in large-N matrix models 
and the consistency conditions of the loop equations yield an algebra which 
is a generalization of the $W_3$ algebra.  Although our loop equations 
contain only the same information as the $W_3$ constraints, they are 
expected to  
provide us with a geometrical setting for constructing $c= 1/2$ SFT. 
Unfortunately the purpose of constructing SFT in sec 6 was not completely 
fulfilled. 
In this section we will construct loop equations for only $J_{\pm}= 
(J_{\pm}^{(1)} + J_{\pm}^{(2)})/2$ by eliminating $\bar{J}_{\pm}  = 
(J_{\pm}^{(1)} - J_{\pm}^{(2)})/2$ from (\ref{loopeq}) and speculate on
possible application of the result to SFT.   

Let us first consider loop equations (\ref{le+1}), (\ref{le+2}) for + loops.  
Because we need loop equations only for loops corresponding to the source 
$J_+$, we rewrite $T_+(l)$ and $W_+(l)$ in terms of $J_+$, $\bar{J}_+$  
and $D_+ \equiv D_+^{(1)}+D_+^{(2)}$, $\bar{D}_+ \equiv D_+^{(1)}-D_+^{(2)}$.
\beq
T_+ = \frac{1}{4} D_+*D_+ - \frac{1}{4} \bar{D}_+* \bar{D}_+ + g (lJ_+)
\triangleleft D_+ + g (l \bar{J}_+) \triangleleft \bar{D}_+,
\eeq
\beqa
W_+ & = & - 3 (D_+* + 2 g (lJ_+) \triangleleft)T_+ + (D_+* +3g(lJ_+) 
\triangleleft )^2 D_+ \nonumber \\
& & - 3g(l \bar{J}_+) \triangleleft (D_+* \bar{D}_+ - g (l \bar{J}_+) 
\triangleleft D_+ + g (lJ_+) \triangleleft \bar{D}_+) \nonumber \\
& & +3g D_+*((l \bar{J}_+) \triangleleft \bar{D}_+) + 3g^2 (lJ_+) \triangleleft
((l \bar{J}_+) \triangleleft \bar{D}_+).
\label{W+}
\eeqa
This result shows that we can eliminate $\bar{D}_+$ from the loop equations 
by setting $\bar{J}_+=0$.  By using (\ref{W+}) and the equation in footnote 4,
we obtain
\beqa
& & l^2 \{ (D_+* + 3g (lJ_+) \triangleleft )^2 D_+ \} (l) Z_+ \nonumber \\
& & = -3gl^2 \frac{\partial}{\partial t} D_+(l) Z_+ + 32 (6\delta''(l) -
t \delta(l) ) Z_+ .
\label{prd}
\eeqa
This agrees with the result in \cite{iikmns}, where the right-hand side in 
(\ref{prd}) was set to zero (denoted $\approx 0$) as being terms either 
with a support at $l=0$ or proportional to backgrounds.

Similarly we can derive loop equations for $J_+$ and $J_-$ from 
(\ref{loopeq}).         
We obtain \footnote{The equation in footnote 6 is also used}
\beqa
& & l^2 \{ D_+ *D_+ *D_+  
 + 3 g D_+ * ((lJ_+) \triangleleft D_+) 
+ 3 g (lJ_+) \triangleleft (D_+ *D_+) \nonumber \\ 
& & - 3 g (lJ_-) *D_+ *D_+ - 3 g D_+ *((lJ_-) \triangleright D_-) 
 - 3 g (lJ_-) \triangleright (D_- *D_-) \nonumber \\ 
& & - 9 g^2 (lJ_-) \triangleright ((lJ_-) \triangleleft D_-) 
+ 9 g^2 (lJ_+) \triangleleft ((lJ_+) \triangleleft D_+) \nonumber \\
& & - 9 g^2 (lJ_-) * ((lJ_+) \triangleleft D_+) \}(l) Z_{+-} \nonumber \\
= &  & 32 \{ 6 \delta''(l) - t \delta (l) \} Z_{+-} 
 - 3 g l^2 D_+(l) \frac{\partial}{\partial t} Z_{+-} 
+ 9 g^2 l^3 J_-(l) \frac{\partial}{\partial t} Z_{+-} 
\label{newprd}
\eeqa 
and the counterpart obtained by interchanging + and $-$ in (\ref{newprd}).
These are generalizations of (\ref{prd}) to + and $-$ loops. 
Let us note that while $X_+(l)$ in (\ref{Xplus}) contain terms which are cubic 
in $J_{\pm}$'s, these terms are canceled altogether in the above equation.    

This loop equation may serve as a starting point for construction of SFT for 
$c= \frac{1}{2}$ string.  We can show that (\ref{newprd}) can be decomposed 
into a set of loop equations.
\beqa
& & \{ D_+*D_+ + 3g (lJ_+) \triangleleft D_+ -3g (lJ_-)*D_+ -3g (lJ_-) 
\triangleright D_- + \frac{\delta}{\delta K_+} \} Z_{+-} 
 \approx 0, \nonumber \\
& & \{ D_+* \frac{\delta}{\delta K_+} +3g (lJ_+) \triangleleft  
\frac{\delta}{\delta K_+} -3g (lJ_-) \triangleright  \frac{\delta}
{\delta K_-} \} Z_{+-} \approx 0
\label{newprd2}
\eeqa
Here $K_+(l)$, $K_-(l)$ are source functions for some auxiliary loop fields.
It is understood that $K_{\pm}(l)$ are set to zero after differentiation in 
(\ref{newprd2}).  These are generalization of the loop equations considered 
in \cite{iikmns}.
\beq
\{ \frac{\delta}{\delta J_0} * 
\frac{\delta}{\delta J_n} + (lJ_0) \triangleleft \frac{\delta}{\delta J_n}
+ \frac{\delta}{\delta J_{n+1}} \} Z |_{J_i=0} \approx 0 \quad (n=0,1,2)
\eeq
Here $J_0$ is identical to $J_+$, while $J_1$ is a  source for a + loop with 
a single operator ${\cal H}$ insertion, and $J_2$ that with double 
${\cal H}$ insertion. Construction of SFT Hamiltonian based on loop 
equation (\ref{newprd2}) is under study.\cite{ns3}

Whether loop equation (\ref{newprd}) is compatible with that of 
Ishibashi and Kawai \cite{ik2}
\beq
l \{ D_+ * D_+ +  D_+ \triangleright D_- + g (lJ_+) \triangleleft D_+ \}(l)  Z
=0
\eeq
is an important problem. This is under investigation and 
we hope to report on this in the future \cite{ns3}.  If these loop equations
are equivalent, the present work will provide us with a connection of 
Ishibashi-Kawai's SFT for $c= 1/2$ gravity to $W_3$ constraints.  
If not, $c=1/2$ SFT has to be constructed from loop equations (\ref{newprd}).

The structures of the loop equations (\ref{loopeq}) are quite interesting.
For example the Virasoro constraint $U_+(l)$ is obtained by 
performing Bogoliubov transformation (\ref{bogo}) on $T_+(l) - V_-(l)$. 
These mathematical structures may worth further investigation. It should also 
be pointed out that the Virasoro generators $U_{\pm}$ involve terms quadratic 
in $J$'s and the $W_3$ currents $X_{\pm}$ contain cubic $J$ terms.

We can extend the present analysis to $c = 1 - \frac{6}{m(m+1)}$ string.  
It is known that this theory can be constructively defined in terms of 
$(m-1)$-matrix-chain model and that the generating function of correlation 
functions in this theory satisfies $W_m$ constraints.  So corresponding to 
the n-th matrix $M_n$ $(n=1,\cdots,m-1)$ we need to introduce $m-1$ source 
functions $J_n^{(r)}(l) \quad (r = 1,\cdots, m-1)$.  It will be 
straightforward to construct loop equations at least for two matter 
configurations on the loops corresponding to the ends $M_1$, $M_{m-1}$ 
of the matrix chain. 

\section*{Acknowledgements}
\hspace{5mm}

One of the authors (R.~N.) thanks J.~Ambj\o rn, N.~Ishibashi , H.~Kawai, 
N.~Kawamoto and C.~Kristjansen for discussions.   

This work is supported in part by Grant-in-Aid for Scientific Research 
(No.07640364) from the Ministry of Education, Science and Culture, 
Japan.

\vspace{1cm}

\appendix

\section{Equivalence of Loop Eq (4.5) to the Virasoro Constraint (2.2)}
\cleqn
\hspace{5mm}
In this appendix we will show that the loop equation (\ref{newle1})
is equivalent to the Virasoro constraint (\ref{v}).

The generating function $Z_{+-}$ can be factorized into two parts.
\beq
Z_{+-} = Z_{+-}^{sing} Z_{+-}^{reg}
\eeq
The singular part $Z_{+-}^{sing}$ is an exponential of quadratic forms of $J$'s
and given by
\beqa
\ln Z_{+-}^{sing} & = &  
 \frac{\sqrt{3}}{2\pi}g  \int_0^{\infty} dl\int_0^{\infty} dl' 
\frac{l^{1/3} l'^{2/3}}
{l+l'} \{ J_+^{(1)}(l) J_+^{(2)}(l') + J_-^{(1)}(l) J_-^{(2)}(l') \}
\nonumber \\
& & -\frac{\sqrt{3}}{2\pi}g  \int_0^{\infty} dl \not\!\!\int_0^{\infty} dl' 
\frac{l^{1/3} l'^{2/3}}
{l-l'} \{ J_+^{(1)}(l) J_-^{(2)}(l') + J_-^{(1)}(l) J_+^{(2)}(l') \} \nonumber 
\\
& & + \frac{2^{4/3}}{3 \Gamma(2/3)} \int_0^{\infty} dl 
(\frac{4}{3} l^{-7/3} - t l^{-1/3}) \{ J_+^{(2)}(l) + J_-^{(2)}(l) \}
\eeqa
Here the symbol $\not\!\int$ means a principal value of an integral.

When $\tilde{D}_{\pm}^{(r)}(\zeta)$ operates on the regular part 
$Z_{+-}^{reg}$, it is assumed to have the following expansions.
\beqa
\tilde{D}_+^{(r)}(\zeta) Z_{+-}^{reg} & = & g \sum_{n=0}^{\infty}
\zeta^{-n-1-r/3} \frac{\partial}{\partial \mu_{3n+r}} Z_{+-}^{reg},
\\
\tilde{D}_-^{(r)}(\zeta) Z_{+-}^{reg} & = &  g \sum_{n=0}^{\infty}
(-1)^{n+r+1}
\zeta^{-n-1-r/3} \frac{\partial}{\partial \lambda_{3n+r}} Z_{+-}^{reg}
\eeqa
Here $\mu_n$ and $\lambda_n$ are source variables for scaling operators and 
defined by
\beqa
\int_0^{\infty} dl J_+^{(r)}(l) l^{n+r/3} & = & 
 \frac{1}{g} \Gamma (n+1+ \frac{r}{3}) \mu_{3n+r}, \\
\int_0^{\infty} dl J_- ^{(r)}(l) l^{n+r/3} & = & 
 \frac{1}{g} \Gamma (n+1+ \frac{r}{3}) \lambda_{3n+r} (-1)^{n+r}
\eeqa

By using these definitions we will rewrite the loop equation (\ref{newle1})
into a differential equation for $Z_{+-}^{reg}$. 
After shifting the source variables according to
\beqa
\mu_1 \rightarrow \mu_1 +  (2)^{1/3} t, & \quad &
  \mu_7 \rightarrow \mu_7 - \frac{3}{7} 2^{1/3}, \nonumber \\
\lambda_1 \rightarrow \lambda_1 + (2)^{1/3} t, & \quad &
  \lambda_7 \rightarrow \lambda_7 - \frac{3}{7} 2^{1/3}, 
\eeqa
we obtain
\beqa
\lefteqn{
\partial_{\zeta} \{ g^2 \sum_{n,m=0}^{\infty} \zeta^{-n-m-3} 
\frac{\partial}{\partial \mu_{3n+1}} \frac{\partial}{\partial \mu_{3m+2}}
} \nonumber \\
& & + g \sum_{r=1,2} \sum_{n=0}^{\infty} \sum_{m=0}^{n+1} \zeta^{m-n-2} 
(m+ \frac{r}{3})(\mu_{3m+r} + \lambda_{3m+r}) \frac{\partial}{\partial 
              \mu_{3n+r}} \nonumber \\
& & + g \sum_{r=1,2} \sum_{n=0}^{\infty} \sum_{m=n+3}^{\infty} \zeta^{m-n-2} 
(m+ \frac{r}{3})\lambda_{3m+r}( \frac{\partial}{\partial 
              \mu_{3n+r}} - \frac{\partial}{\partial 
              \lambda_{3n+r}})                      \nonumber \\
& & + \frac{g}{9} \frac{1}{\zeta^2} + \frac{2}{9} \frac{1}{\zeta} 
(\mu_1 + \lambda_1)(\mu_2 + \lambda_2) \} Z_{+-}^{reg} = 0
\label{a8}
\eeqa
Let us note that the third term is composed of positive powers of $\zeta$. 
This term, however, shows  that $Z_{+-}^{ reg}$ depends on 
$\mu$ and $\lambda$ only through the combination $\mu_n + \lambda_n$.    
Then (\ref{a8}) reduces to the Virasoro constraint (\ref{v}) on 
$\tau (\mu + \lambda) = Z_{+-}^{reg}$.

\section{`Extended' $W_3$ Algebra}
\cleqn
\hspace{5mm}
The operator $X_+(l)$ defined in sec 5 is given by 
\beqa
X_+(l) & = & \sum_{r=1}^{2} \{ D_+^{(r)} *D_+^{(r)} *D_+^{(r)} + 3 g
(lJ_+^{(3-r)}) \triangleleft (D_+^{(r)} *D_+^{(r)}) \nonumber \\ 
& & - \frac{3}{2} g (lJ_-^{(3-r)}) *D_+^{(r)} *D_+^{(r)} 
- 3 g ((lJ_-^{(r)}) \triangleright D_-^{(3-r)}) \triangleright D_-^{(3-r)} 
\nonumber \\ 
& & + 3 g^2 (lJ_+^{(3-r)}) \triangleleft ((lJ_+^{(3-r)}) \triangleleft 
D_+^{(r)} ) + \frac{3}{4} g^2 (lJ_-^{(3-r)}) *(lJ_-^{(3-r)}) *D_+^{(r)} 
\nonumber \\ 
& & - 3 g^2 (lJ_+^{(3-r)}) \triangleleft ((lJ_-^{(3-r)}) * D_+^{(r)}) 
+ 3 g^2 ((lJ_-^{(3-r)}) \triangleright (lJ_+^{(3-r)})) \triangleright 
D_-^{(r)} \nonumber \\ 
& & - 3 g^2 ((lJ_-^{(r)}) *(lJ_-^{(r)})) \triangleright D_-^{(3-r)}  
 - \frac{9}{8}g^3 (lJ_-^{(r)}) *(lJ_-^{(r)}) *(lJ_-^{(r)}) \nonumber \\ 
& & - \frac{9}{4} g^3 (lJ_+^{(r)}) \triangleleft ((lJ_+^{(r)})  
\triangleleft (lJ_-^{(r)})) 
+ \frac{9}{4} g^3 (lJ_+^{(r)}) \triangleleft ((lJ_-^{(r)})*(lJ_-^{(r)})) \}(l) 
\nonumber \\
& & 
\label{Xplus}
\eeqa

In the remaining part of this Appendix we will show the algebra 
that $U_{\pm}$ and $X_{\pm}$ generate.
First of all $T_{\pm}$, $V_{\pm}$, $W_{\pm}$ and $Y_{\pm}$ generate the 
following algebra.  
Some formulae which are not presented here can be found in (\ref{cm1})$\sim$
(\ref{cm3}), (\ref{cm4}), (\ref{cm5}). The index $\pm$ will be suppressed here.
\beq
[T(l), Y(l')]  =  g (2l + l') W(l-l') \theta(l-l') 
   + g (2 l + l') Y(l'-l) \theta(l'-l),
\eeq
\beq
[V(l), W(l')] = - g (2l+l') Y(l-l') \theta(l-l') 
 - 2 ( 2 l' + l) W(l'-l) \theta(l'-l), 
\eeq
\beq
[V(l), Y(l')] = - g (2l-l') Y(l+l'),
\eeq
\beqa
[Y(l), Y(l')] & = & -9 g (l-l') (V*V)(l+l') -18g(l-l') (V \triangleright T)
(l+l') \nonumber \\
& & + \frac{3}{2} g^2 (l-l') (l^2 + 4 ll' + l'^2) V(l+l'), 
\eeqa
\beqa
 [Y(l), W(l')] &  & = \theta(l'-l) \{ -9g (l+l') (T*T)(l'-l)  \nonumber \\
 & & -18 g (l+l') 
(V \triangleleft T)(l'-l) 
+ \frac{3}{2} g^2 (l+l') (l^2-4ll'+l'^2) T(l'-l) \}
\nonumber \\
& & + \theta(l-l') \{ -9g (l+l') (V*V)(l-l') -18 g (l+l') (V \triangleright T)
(l-l') \nonumber \\
& & + \frac{3}{2} g^2 (l+l') (l^2-4ll'+l'^2) V(l-l') \}
\eeqa

By using these relations we can derive the algebra of 
$\hat{U}_{\pm}  =  T_{\pm} - V_{\mp}$ and $\hat{X}_{\pm}  =  
W_{\pm} - Y_{\mp}$.
\beq
[ \hat{U}_{\pm}(l), \hat{X}_{\pm}(l')] = g (2l-l') \hat{X}_{\pm}(l+l'),
\eeq
\beq
[ \hat{U}_{\pm}(l), \hat{X}_{\mp}(l') ]   = -  g (2l+l') \hat{X}_{\pm}
(l-l') \theta(l-l') + g (2 l + l') \hat{X}_{\mp}(l'-l) \theta(l'-l), 
\eeq
\beqa
[\hat{X}_{\pm}(l), \hat{X}_{\pm}(l')] & = & 
-\frac{3}{2} g^2 (l-l') (l^2 + 4 ll' + l'^2) \hat{U}_{\pm}(l+l') \nonumber \\
& & + 9g(l-l') \{ T_{\pm}*\hat{U}_{\pm} + V_{\mp}*\hat{U}_{\pm}  
+ 2 V_{\pm} \triangleleft \hat{U}_{\pm} \nonumber \\
& & - 2 V_{\mp} \triangleright \hat{U}_{\mp} \} (l+l'),
\eeqa
\beqa
[\hat{X}_+(l),\hat{X}_-(l')] & = & \theta(l-l') [ \frac{3}{2} g^2(l+l')
(l^2-4ll'+l'^2) \hat{U}_+(l-l') \nonumber \\ 
& & -9g(l+l') \{ T_+ * \hat{U}_+ 
+ V_- * \hat{U}_+ + 2 V_+ \triangleleft \hat{U}_+ \nonumber \\ 
& & -2V_- \triangleright \hat{U}_- \} (l-l') ] 
\nonumber \\ 
& & -(l \leftrightarrow l' ,+ \leftrightarrow -)
\eeqa
Other commutation relations are given in (\ref{con1}) and (\ref{con2}).  
The algebra of $U_{\pm}$ and $X_{\pm}$ can be derived from the above by 
transformation (\ref{bogo}).


\newpage

\begin{thebibliography}{99}

\bibitem{kazakov}
V. Kazakov, Mod. Phys. Lett. {\bf A4} (1989) 2125.

\bibitem{fkn}
M. Fukuma, H. Kawai and R. Nakayama, 
Int. J. Mod. Phys. {\bf A6} (1991) 1385.

\bibitem{dvv}
R. Dijkgraaf, E. Verlinde and H. Verlinde, 
Nucl. Phys. {\bf B348} (1991) 435. 

\bibitem{kostovnokori}
I. K. Kostov, Nucl. Phys. {\bf B326} (1989) 583; {\em ibid.} {\bf B376} 
(1992) 539;Phys. Lett. {\bf B266} (1991) 42;I. Kostov and M. Staudacher, 
Nucl. Phys. {\bf B384} (1992) 459; 
B.Eynard and C.Kristjansen, Nucl. Phys. {\bf B455} (1995) 577.

\bibitem{kaku}
M.Kaku and K.Kikkawa, Phys. Rev. {\bf D10} (1974) 1110, 1823;
W.Siegel, Phys. Lett. {\bf B151} (1985) 391, 396; 
E.Witten, Nucl. Phys. {\bf B268} (1986) 253; 
Hata, K.Itoh, T.Kugo, H.Kunitomo and K.Ogawa, Phys. Lett. {\bf B172} (1986)
186, 195; 
A.Neveu and P.West, Phys. Lett. {\bf B168} (1986) 192; 
M.Kaku, Int. J. Mod. Phys. {\bf A9} (1994) 139.

\bibitem{ik}
N. Ishibashi and H. Kawai, Phys. Lett. {\bf B314} (1993) 190. 

\bibitem{ik2}
N.~Ishibashi and H.~Kawai, Phys. Lett. {\bf B322} (1994) 67.

\bibitem{iikmns}
M.~Ikehara, N.~Ishibashi, H.~Kawai, T.~Mogami, R.~Nakayama and N.~Sasakura,
Phys. Rev. {\bf D50} (1994) 7467; 
{\it A Note on String Field Theory in the Temporal Gauge},
proceedings of the Workshop on {\it Quantum Field Thoery, Integrable Models 
and Beyond}, Yukawa Institute for Theoretical Physics, Kyoto University, 
14-18 Feb 1994, Prog. Theor. Phys. Supp. {\bf 118} (1995) 241.

\bibitem{ns}
R. Nakayama and T. Suzuki,
Phys. Lett. {\bf B354} (1995) 69.

\bibitem{wata}
Y.Watabiki, Nucl. Phys. {\bf B441} (1995) 119; Phys. Lett. {\bf 346} (1995)46.

\bibitem{ike}
M.~Ikehara, Phys. Lett {\bf B348} (1995) 365; Prog. Theor. Phys. {\bf 93}
(1995) 1141.

\bibitem{jr}
A.Jevicki and J.Rodrigues, Nucl. Phys. {\bf B421} (1994) 278.

\bibitem{sy}
F. Sugino and T. Yoneya, {\it Stochastic Hamiltonian for Noncritical String 
Field Theories}, UT-Komaba-95-8, Oct 1995.

\bibitem{gava}
E.Gava and K.S.Narain, Phys. Lett. {\bf B263} (1991) 213.

\bibitem{kkm}
V.A. Kazakov, I.K. Kostov and A.A. Migdal, Phys. Lett. {\bf B157} (1985)
295;
J. Ambj{\o}rn, B. Durhuus and J. Fr\"{o}hlich, Nucl. Phys. {\bf B257}
[FS14] (1985) 433;
F. David, Nucl. Phys. {\bf B257} [FS14] (1985) 543.

\bibitem{kaza}
V.A.Kazakov, Phys. Lett. {\bf A119} (1986) 140; 
D.V.Boulatov and V.A.Kazakov, Phys. Lett. {\bf B186} (1987) 379.

\bibitem{matrix}
E. Br\'{e}zin and V. Kazakov, Phys. Lett. {\bf B236} (1990) 144;
M. Douglas and S. Shenker, Nucl. Phys. {\bf B335} (1990) 635;
D. J. Gross and A. Migdal, Phys. Rev. Lett. {\bf 64} (1990) 127;Nucl. Phys.
{\bf B340} (1990) 333.

\bibitem{bdss}
T.~Banks, M.~Douglas, N.~Seiberg and S.~Shenker, Phys. Lett. {\bf B238} (1990)
43.

\bibitem{ns2}
R.~Nakayama and T.~Suzuki, {\it New Loop Equations in Ising Model Coupled to
2d Gravity and String Field Theory}, in the Proceedings of the Workshop held
in honor of the 60th birthday of Professor Keiji Kikkawa ``Frontiers in 
Quantum Field Theory'', Osaka, Japan, Dec 1995.

\bibitem{zamo}
A.B. Zamolodchikov, Theor. Math. Phys. {\bf 65} (1985) 1205.

\bibitem{kostov}
I. Kostov, Phys. Lett. {\bf B266} (1991) 317.

\bibitem{mss}
G. Moore, N. Seiberg and M. Staudacher, Nucl. Phys. {\bf B362} (1991) 665.

\bibitem{dks}
J.-M.~Daul, V.A.~Kazakov and I.K.~Kostov, Nucl. Phys. {\bf B409} (1993) 311 ;
M.~Anazawa, A.~Ishikawa and H.~Itoyama, Phys. Lett. {\bf B362} (1995) 59;
M.~Anazawa and H.~Itoyama, {\it Macroscopic N Loop Amplitude for Minimal 
Models Coupled to Two-Dimensional Gravity}, preprint OU-HET-222 (Nov 1995).

\bibitem{aw2}
J. Ambj\o rn and Y. Watabiki, {\it Non-critical string field theory for 2d
quantum gravity coupled to (p,q)-conformal fields}, NBI and TIT preprint
(1996).

\bibitem{kkmw}
H. Kawai, N. Kawamoto, T. Mogami and Y. Watabiki, Phys. Lett. {\bf 306B} (1993)
19; 
S.S.~Gubser and I.R.~Klebanov, Nucl. Phys. {\bf B416} (1994) 827; 
J.~Ambj\o rn and Y.~Watabiki, Nucl. Phys. {\bf B445} (1995) 129.

\bibitem{ajw}
J.~Ambj\o rn, J.~Jurkiewicz and Y.~Watabiki, Nucl. Phys. {\bf B454} 313;
S.~Catteral, G.~Thorleifsson, M.~Bowick and V.~John, 
Phys. Lett. {\bf B354} (1995) 58.

\bibitem{ns3} 
R.~Nakayama and T.~Suzuki, in preparation.

\end{thebibliography}
\end{document}